\documentstyle[12pt,fleqn,cite]{article}
\def\mathbf#1{\mbox{\boldmath $#1$}}
\newfont\myb{msbm8} 
\def\b#1{\hbox{\myb#1}}
\newfont\mybb{msbm10}
\def\bb#1{\hbox{\mybb#1}}
\def\C {\bb{C}}
\def\c {\b{C}}
\def\R {\bb{R}}

\def\prl#1#2#3{{\it Phys. Rev. Lett.}{\bf #1}({#2}){#3}}

\def\jmp#1#2#3{{\it J. Math Phys.} {\bf #1}({#2})#3}

\def\jpa#1#2#3{{\it J. Phys.} {\bf A #1}({#2})#3}
\def\phrep#1#2#3{{\it Phys. Rep.}{\bf {#1}}({#2}){#3}}

\def\ibid#1#2#3{{\it ibid.}{\bf { #1}}({#2}){#3}}
\long\def\cf#1#2#3{\bibitem{#1} {#2}, {#3}.}
\def\eq#1{(\ref{#1})}
\def\eqs#1#2{(\ref{#1}) and (\ref{#2} )}
\def\be{\begin{equation}} 
\def\ee{\end{equation}}
\def\bea{\begin{eqnarray}} 
\def\eea{\end{eqnarray}}

\def\rt{\longrightarrow} 
\def\papertitle#1{\title{\vspace{-2cm}
\flushright{\small IP/BBSR/ 96-46 \\ hep-th/9609025} \\
\vspace{2cm} \begin{center} {#1} \end{center}}}
\begin{document} 
%\begin{titlepage}
\papertitle{A Quantum Many-body Problem in Two Dimensions: 
Ground State} 
\author{Avinash Khare\thanks{khare@iopb.ernet.in} 
and Koushik Ray\thanks{koushik@iopb.ernet.in} \\
Institute of Physics, \\Bhubaneswar 751 005, INDIA} 
\date{}
\maketitle
\begin{abstract}
\noindent We obtain the exact ground state for the Calogero-Sutherland 
problem in arbitrary dimensions. 
In the special case of  two dimensions, we show that the problem 
is connected 
to the random matrix problem for
complex matrices, provided the strength of the inverse-square interaction 
$g = 2$. In the thermodynamic limit, we obtain the ground state energy 
and the
pair-correlation function and show that in this case there is no 
long-range order.
\end{abstract}
\thispagestyle{empty}
%\end{titlepage}

\noindent Calogero-Sutherland type models are a class of
exactly solvable models. Starting from the inception\cite{calo}, 
till today,
these models continue to be of interest, especially due to their 
exact solvability
\cite{pere}. Moreover, these models are related to (1 + 1)-dimensional 
conformal
field theory, random matrices and a host of other things\cite{sim}.
Some variants of the
models have attracted attention in very recent past\cite{sim,khare,pkg}.  
The original model of Calogero\cite{calo} described $N$ particles in 
one dimension, 
interacting through an inverse square law and a two-body harmonic 
interaction. 
The latter interaction was added to the effect 
of complete discretization of the spectrum. 
This model gave in to an analytic treatment yielding an exact solution. 
Sutherland\cite{sutherland} considered a variant of
this problem where the harmonic interaction was replaced by a harmonic 
well  
containing the $N$ particles. 
It was shown that one can exactly solve for the ground state of the 
system. The norm
of the ground state wave-function was shown to be of a  form that 
coincides with the
joint probability density function of the eigenvalues of random 
matrices from an orthogonal, unitary or symplectic Gaussian
ensemble, provided, the parameter giving the strength of the 
inverse-square 
interaction was fixed to $\frac{1}{2}$, 1 or  2 respectively.
In an effort to generalize the original Calogero model\cite{calo} to 
dimensions higher than one, it was shown 
that some exact eigenstates including the ground state can be 
obtained
for a three dimensional $N$-body problem with  inverse-square 
interaction, 
provided one also adds a long-range three-body interaction\cite{cm}, 
which is not 
present in one dimension. 
It is clearly of interest to enquire if this result of \cite{cm}
could be generalized to arbitrary dimensions. Perhaps a much more 
interesting
question is whether one could obtain the exact ground state for the 
Sutherland
variant of this model in arbitrary dimensions. One could then hope that 
a la one
dimensional case, one could map the problem to some random matrix 
problem and obtain
exact results for the corresponding many-body problem. 
The purpose of this letter is to provide one such example. 
In particular, 
for the special case of two dimensions, we show that the norm of the 
ground
state wave-function is related to  the  joint probability density 
function of the eigenvalues of complex random matrices, 
provided, the strength of the inverse square interaction$(g)$ is 
set  to  2. 
Using this correspondence, we obtain some exact results for the 
ground state of a
two-dimensional many-body system in the thermodynamic limit. In 
particular, in the
limit $ N \rt \infty$, $\omega \rt 0$ with $N\omega $ fixed, we 
obtain the $n$-point
correlation function. Explicit expressions for one- and two-point 
correlation
functions are also given and it is pointed out that in this case 
($ g = 2$), 
unlike in one dimension, the particle
density does not have the semi-circular form and the system does 
not have any
long-range order. 

Let us start by describing the model in $D$ dimensions. 
Following \cite{cm}, we shall first consider the $N$-body problem 
in $D$ dimensions, defined by the Hamiltonian 
describing $N$ particles interacting through an inverse square potential, 
a harmonic potential and a three body term: 
\bea \label{hamilcm}
{\cal H} = -\sum_{i=1}^{N} \nabla_i^2 + g \!\!\! \sum_{\{i,j|i<j\}}^N 
\frac{1}{r_{ij}^2} 
+ \quad G \!  \!\!\!\!\! \sum_{\stackrel{\{i,j,k \}}{i<j, i \ne k, 
j \ne k} }
\frac{{\bf r}_{ki}{\bf . r}_{kj} }{r_{ki}^2 r_{kj}^2} + 
\frac{1}{8}\omega^2 \sum_{i=1}^N r_{ij}^2,  
\eea 
where we  have chosen units such that $ \hbar^2 = 2m = 1$. 
Following \cite{cm}, it is easily shown that some eigenstates 
including the ground
state for the system are given by, 
\bea
\Psi_n = \left( \prod_{i < j} r_{ij}^2 \right)^\frac{\Lambda_D}{2} 
\exp \left( - \frac{\omega}{4} \sqrt{\frac{1}{2N}}\sum_{\{i,j| i< j\}} 
r_{ij}^2 \right)
L_n^{\Gamma_D}\left( \frac{\omega}{2} \sqrt{\frac{1}{2N} } 
\sum_{\{i,j| i< j\}} r_{ij}^2 \right)
\eea
with energy 
\bea
E_n = \sqrt{\frac{N}{2}}\left( 2n + \Gamma_D +1 \right) \omega.
\eea
Here ${\bf r}_i$ is the $D$-dimensional position vector of the $i$-th 
particle and ${\bf r}_{ij} = {\bf r}_i - {\bf r}_j $ denotes the 
relative separation of 
the $i$-th and $j$-th particles, while $r_{ij}$ denotes its magnitude. 
The parameters $g$ and $G$ are the strengths of the 
two- and three-body potentials, respectively. 
Here $g > -1/2$ to stop the {\em fall to the origin}.  
Further, $\Lambda_D$  and $\Gamma_D$ are two 
parameters determined in terms of the parameters of the 
Hamiltonian by, 
\bea\label{binG}
\Lambda_D & \equiv & \sqrt{\frac{G}{2}} \\ \label{bing}
& = & \frac{1}{2} \left[ \sqrt{(D -2)^2 + 2g} - (D -2 ) \right] \\ 
\label{Gamma}
\Gamma_D & = & \frac{1}{2} \left[ D (N - 1) - 2 + \Lambda_D N ( N -1) 
\right],
\eea
where $L_n^{\Gamma_D} $ denotes the Laguerre polynomial. 
As expected, the results are in agreement with \cite{calo} and 
\cite{cm} for $D=1 $
and   3  respectively.
It is worth noting here that the case $D=2$ is somewhat special, 
since the results
are valid only for $g \ge 0 $. Further, for any $D > 1$,  as $G$ 
and $g \rt 0 $, 
the $n =0$ state corresponds to the ground state of the oscillator 
problem without
the centrifugal barrier and with Bose statistics. 
Thus the situation is
different from the case in one dimension\cite{sutherland}, where, 
as $ g \rt 0$, the 
ground state is that of  the oscillator problem without the 
centrifugal barrier 
and  with Fermi statistics.
It is not clear whether the difference  in statistics in the cases 
of one and more
than one dimensions has any deeper physical significance.

Now let us consider a Sutherland variant of the above problem, 
where now the system
is described by the Hamiltonian,
\bea \label{hamil}
{\cal H} = -\sum_{i=1}^{N} \nabla_i^2 + g \!\!\! \sum_{\{i,j|i<j\}}^N 
\frac{1}{r_{ij}^2} 
+ \quad G \!  \!\!\!\!\! \sum_{\stackrel{\{i,j,k \}}{i<j, i 
\ne k, j \ne k} }
\frac{{\bf r}_{ki}{\bf . r}_{kj} }{r_{ki}^2 r_{kj}^2} + 
\omega^2 \sum_{i=1}^N r_i^2.  
\eea 
This Hamiltonian is obtained by replacing the $\frac{1}{8}
\omega^2\sum_{i>j=1}^N
r_{ij}^2$ in \eq{hamilcm} with the harmonic well potential 
$\omega^2 
\sum_{i=1}^N r_i^2$ in the same spirit as in \cite{sutherland}. 

The ground state eigenfunctions and eigenvalues of the system\eq{hamil} 
are given by,
\bea\label{grst}
\Psi_0 &=& \hat{C}\exp\left( -\frac{\omega}{2} \sum_{i=1}^N r_i^2 \right)
 \prod_{i<j} \mid {\bf r}_i - {\bf r}_j \mid^{\Lambda_D},   \\
\label{gren}
E_0 &=& \left[ DN + N(N - 1) \Lambda_D \right] \omega,
\eea
where $\Lambda_D $ is as given in \eqs{binG}{bing}.
As in the previous case, as $G$ and $g \rt 0$, the ground
state becomes that for non-interacting particles in a harmonic potential, 
without the centrifugal barrier and with Bose statistics. 

Now, rewriting $\Psi_0$ from \eq{grst} in terms of a new variable
\bea
{\bf y}_i \equiv \sqrt{\frac{\omega}{\Lambda_D}}{\bf r}_i,
\eea
we have the probability distribution for the $N$ particles as
\bea\label{psi2}
\Psi_0^2 = C \exp \left( - \Lambda_D \sum_{i=1}^N y_i^2
\right)\prod_{i<j} \mid {\bf y}_i - {\bf y}_j \mid^{2\Lambda_D},
\eea
where $C$ is the constant of  normalization.

In the case of $D=1$, there is no three-body interaction\cite{cm},
{\em i.e.} $G = 0$ and  $\Lambda_1$ is 
determined from the definition\eq{bing}. It was shown in 
\cite{sutherland} that 
for $\Lambda_1 = \frac{1}{2}, $ 1 and 2, 
the corresponding expressions for $\Psi_0^2$ match with the 
distribution functions 
for the eigenvalues of random matrices from a Gaussian ensemble.
In particular, they correspond respectively to the
orthogonal, unitary and symplectic matrices\cite{mehta}. 
Using the well-known results derived earlier\cite{mehta}, 
Sutherland \cite{sutherland}
wrote down the expressions for
density and pair-correlation functions for a many-body system 
in one dimension.

Our concern for the rest of this letter will be to deal with 
the case of $D=2$. We shall
show that in this case also the expression for $\Psi_0^2$ renders 
itself to 
identification with the distribution of eigenvalues from an ensemble 
of complex
matrices, provided, $g = 2$. 

In two-dimensions, the ground state energy\eq{gren} becomes
\bea
E_0 = [2N + N ( N - 1 ) \Lambda_2 ] \omega,
\eea
while the parameter $\Lambda_2$ becomes 
\bea
\Lambda_2  =  \sqrt{\frac{G}{2}} 
 = \sqrt{\frac{g}{2}},
\eea
so that the strengths of both the 
interactions are equal ($G = g$). The form of the distribution \eq{psi2} 
remains the same, with ${\bf y}_i$ denoting two-dimensional vectors.
Now, one can write \`two-vectors' as complex numbers, the 
two-dimensional space of
positions of the particles becoming the complex plane, $\C$. 
Let us denote the complex
numbers corresponding to the two-vectors ${\bf y}_i \in \R^2$  
by $z_i \in \C$, 
and rewrite \eq{psi2} in terms of $z_i$ as, 
\bea\label{psiz}
\left[\Psi_0(z_i)\right]^2 = C \exp \left( - \Lambda_2 \sum_{i=1}^N 
\mid z_i\mid^2
\right)\prod_{i<j} \mid z_i - z_j \mid^{2\Lambda_2},
\eea
The form \eq{psiz} of $\Psi_0^2$ is known to occur in the study of 
complex random
matrices. This is precisely the form of the joint probability 
density function of the eigenvalues of matrices from 
an ensemble of complex matrices\cite{mehta,ginibre}, 
provided one sets $\Lambda_2$ equal to 1. The normalization constant 
can be determined as in
\cite{mehta,ginibre}, and the expression for $\Psi_0^2$ with 
$\Lambda_2 =1$  and 
normalized to unity is 
\bea\label{psin}
[\Psi_0(z_i)]^2 = \left(\pi^N \prod_{p=1}^N p !\right)^{-1} 
\exp \left( -\sum_{i=1}^N \mid z_i \mid^2
\right)\prod_{i<j} \mid z_i - z_j \mid^{2}.
\eea
It is worth noting here that by fixing $\Lambda_2$, 
one is left with one single dimensional variable in the problem, {\em viz.} 
$\omega$, which has the dimension of $\frac{1}{\rm length^2}$ 
(recall that we are 
working  in the units $\hbar^2 = 2m = 1$). Therefore, 
the thermodynamic limit 
of taking the area and $N$ to be large with their ratio kept 
finite, will be 
achieved by taking $N \rt \infty$, $\omega \rt 0$, with $N\omega = 
\rm constant$.  
We shall show below how to take this singular limit.

Following \cite{mehta,ginibre} one can also find out the $n$-point 
correlation functions 
for all $n$. Let us quote the general result.
The $n$-point correlation function is defined as
\bea
{\cal R}_n(z_1, \cdots , z_n) \equiv \frac{N!}{(N - n )!}\int \ldots \int
[\Psi_0(z_i)]^2 \prod_{i = n+11}^N \mid dz_i\mid^2
\eea
For the case at hand one finds, after calculating the Van-der 
Monde determinant,that
the expression for the $n$-point correlation function is 
\bea\label{npt}
{\cal R}_n(z_1, \cdots , z_n) = \frac{1}{\pi^n} \exp \left( - 
\sum_1^n \mid z_i \mid^2
\right) \det \left[ \sum_{p=0}^{N -1} \frac{(z_iz_j^*)^p}{p!}
\right]_{\{ i,j = 1,2, \cdots n\}}.
\eea
As $N \rt \infty$, the correlation functions tend to well-defined limits:
\bea
{\cal R}_n(z_1, \cdots , z_n) \sim \frac{1}{\pi^n} \exp \left( - \sum_1^n 
\mid z_i \mid^2
\right) \det \left[ e^{z_i z_j^*} \right]_{\{ i,j = 1,2, \cdots n\}}.
\eea
In particular, the one-point correlation function, defined as,
\bea
{\cal R}(z) = N \int_{\c^{N-1}} \Psi_0^2 dz_2 dz_3 \cdots dz_N,   
\eea
and interpreted as the density of the $N$-particle system, 
is given by:
\bea \label{density}
{\cal R}(\zeta) = \frac{1}{\pi} \exp \left( -\zeta^2 \right) \sum_{p=0}^{N-1}
\frac{\zeta^{2p}}{p!}
\eea
where we have put $ \mid z \mid = \zeta $ and omitted the suffix 1:
 ${\cal R} \equiv
{\cal R}_1$.
The density\eq{density} is isotropic and does not depend on  the 
angular coordinate
$\theta \equiv \arg (z) $.
The density is normalized as 
\bea\label{normalizn}
\int_0^{2\pi}\int_0^\infty {\cal R}(\zeta) \zeta d\zeta d\theta = N,
\eea
the total number of particles. It is interesting to note the 
difference between
the density \eq{density} and the semi-circular expression for 
density obtained in
\cite{sutherland} for the corresponding one-dimensional system, {\em viz.} 
\bea \nonumber
{\cal R}(y) & = & \sqrt{2N - y^2}, \qquad y^2 < N, \\ \label{suden} 
& = & 0, \qquad \qquad \qquad y^2 > N.
\eea
Each integral in the sum  in ${\cal R}(\zeta)$ gives a value of unity 
and the $N$ terms of the sum adds upto $N$. This behavior is 
different from the simpler 
form of density in \eq{suden}, where one could factor out 

$\sqrt{N}$ from the 
density. This difference in the form of the density will be 
manifested in the  
different way of achieving the thermodynamic limits in the two cases. 
Let us now write down the expression for the density in terms of the original
position vectors ${\bf r}$. Demanding the same normalization 
as in \eq{normalizn}, this leads to the following expression for density
\bea\label{densr}
{\cal R}(r) = \frac{\omega}{\pi\Lambda_2} \exp \left( -\omega r^2 \right) 
\sum_{p=0}^{N-1}
\frac{(\omega r^2)^{p}}{p!}
\eea

As noted earlier, the thermodynamic limit of the expressions is 
given by $N \rt
\infty$ and $\omega \rt 0$, with $N\omega$ kept constant.  
However, one can see from the expression\eq{densr} that it is not 
meaningful to take the limit $\omega \rt 0$, since that will 
make the density
vanishing. In order to take this limit meaningfully, 
one has to express the density in terms of a
new variable, defined by pulling out a factor of $\sqrt{N}$ 
from ${\bf r}$, 
\bea
{\bf r} = \sqrt{N} \mathbf{\rho}.
\eea
The density ${\cal R}(\rho)$ with the same normalization takes the form:
\bea
{\cal R}(\rho) = \frac{\omega N}{\pi} \exp \left( -N\omega \rho^2 \right) 
\sum_{p=0}^{N-1}
\frac{(N\omega\rho^2)^p}{p!}.
\eea
Denoting the density at $\rho =0$ by ${\cal R}_0$, we find that 
${\cal R}_0 = 
 \frac{\omega N}{\pi} $. Thus, we see that 
for a fixed ${\cal R}_0$, letting  $N \rt \infty$ means $\omega \rt 0$ 
as
$\frac{1}{N}$. 

The pair-correlation function for the system has also been 
obtained\cite{mehta,ginibre}. It is given by
\bea\label{pair}
{\cal R}_2(\xi) = {\cal R}_0^2  \left[ 1 - \exp \left( - \pi\xi^2 \right)
 \right],
\eea 
where  $\xi = \frac{\sqrt{N\omega}}{\pi}\rho_{ij}$.
Let us note at this point that, the pair-correlation function, as given
above in \eq{pair}, has apparent similarity with the pair-correlation 
function for the corresponding one dimensional case\cite{sutherland}, with
 $\Lambda_1 = \frac{1}{2}$, which corresponds to the case of orthogonal 
Gaussian ensemble.
In  both the cases the function ${\cal R}$ starts off from a value of
0 near $\xi = 0$ and grows up to unity as $\xi \rt \infty$.
To facilitate discussion let us compare the functions $Y_2 = 1 - {\cal R}_2$ 
setting  ${\cal R}_0 =1$ in both the cases.  
The behavior of $Y_2$ for our case is 
\bea
Y_2(\xi) = 1 - \pi\xi^2 + \frac{\pi^2\xi^4}{2} - \frac{\pi^3\xi^6}{6} + 
\cdots,
\eea
while for the orthogonal $(\Lambda_1 = \frac{1}{2})$ case of 
\cite{sutherland}, one had
\bea
Y_2(\xi) = 1 - \frac{\pi^2\xi}{6} + \frac{\pi^3 \xi^3}{60} - 
\frac{\pi^4\xi^4}{135}
+ \cdots.
\eea
That is, while for small $\xi$, $Y_2$ goes in a power series 
with only even power of
$\xi$ in our case, for the $\Lambda_1 =\frac{1}{2}$ case in 
one dimension, 
it is a power series in all powers of $\xi$. 
However, for large distances, while the $Y_2$ in \cite{sutherland} 
goes as powers of 
$\frac{1}{\xi^2}$, {\em i.e.}
\bea
Y_2(\xi) = \frac{1}{\pi^2\xi^2} - \frac{1 + \cos^2\pi\xi}{\pi^4\xi^4} 
+ \cdots ,
\eea
with extremely mild oscillations,  
we have a Gaussian fall-off:  $Y_2 = \exp \left( - \xi^2 \right)$, 

without 
any oscillations. 
Moreover, since the Fourier transform of a Gaussian function is a 
Gaussian only, the form factor
for our system is a Gaussian, unlike that in \cite{sutherland}. 

One can also write down the higher order correlations\cite{janco}. 
For example, the three-particle correlation function in the 
variables $\mathbf{\rho}$
is given by:
\bea\label{three}
&&{\cal R}_3(\mathbf{\rho_1}, \mathbf{\rho_2}, \mathbf{\rho_3})
/{\cal R}_0^3
 =  1  - \exp \left( - \pi {\cal R}_0 \rho_{12}^2 \right) 
 -  \exp \left( - \pi {\cal R}_0 \rho_{23}^2 \right) 
 - \exp \left( - \pi {\cal R}_0 \rho_{31}^2 \right) \nonumber \\
& + & 2 \exp \left[ 
-\frac{1}{2} \pi {\cal R}_0 \left( \rho_{12}^2 + \rho_{23}^2 
+ \rho_{31}^2 \right)
\right] \cos \left[ 2\pi {\cal R}_0 {\cal A}(1, 2, 3) \right],
\eea
where ${\cal A}(1, 2, 3)$ is the area of the triangle formed 
by the three particles
1, 2 and 3. 
The four-body correlation function is 
\bea\label{four}
&& {\cal R}_4(\mathbf{\rho_1}, \mathbf{\rho_2}, \mathbf{\rho_3} 
\mathbf{\rho}_4) / {\cal R}_0^4 = 1  
 - \exp \left( - \pi {\cal R}_0 \rho_{12}^2 \right) \cdots 
\nonumber \\
& + &\exp \left[ - \pi {\cal R}_0 \left(\rho_{12}^2 + 
\rho_{34}^2 \right) \right] +
\cdots \nonumber \\
& + & 2 \exp \left[ 
-\frac{1}{2} \pi {\cal R}_0 \left( \rho_{12}^2 + \rho_{23}^2 + 
\rho_{31}^2 \right)
\right] \cos \left[ 2\pi {\cal R}_0 {\cal A}(1, 2, 3) \right] + 
\cdots \nonumber \\
& - & 2 \exp \left[ -\frac{1}{2} \pi {\cal R}_0 \left(\rho_{12}^2 
+ \rho_{23}^2 + 
\rho_{34}^2 + \rho_{41}^2 \right) \right] \cos \left[ 2 \pi{\cal R}_0 
{\cal A}(1, 2, 3, 4) \right],
\eea
where ${\cal A}(1, 2, 3, 4) \equiv \frac{1}{2} 
\mid \mathbf{\rho_{13}}{\bf \times}
\mathbf{\rho_{24}} \mid $ and $\cdots$ stands for permutations 
of the indices.
It  is obvious from \eqs{three}{four} as well as from the
general expression\eq{npt} for the $n$-point correlation functions
that all the distribution functions are isotropic, {\em i.e.}, 
do not depend on the
angular coordinate. Further, they do not show either long-range 
or quasi-long-range order and have a Gaussian fall-off at large distances. 

The form of the distribution function \eq{psi2} is the same\cite{janco}
 as that occurs in the case of
two-dimensional Coulomb gas\cite{dyson,wigner}, although the 
physics in the two cases are quite different.

To conclude, in this letter, we have studied 
a many-body problem in two dimensions. We have written down 
explicit expressions for
the ground state energy and the pair-correlation function, 
in case the strength of the 
inverse square interaction  
is set to $g = 2$. We then show that in this case, there is 
no long-range or
quasi-long-range order. It would be very interesting if one
could obtain exact results for other values of $g$ and see if there is 
long-range order for any value of $g$. 
It will also be interesting 
if one could solve a similar many-body problem in higher dimensions, 
especially in three dimensions. But to our knowledge,
the corresponding random matrix problem has not been worked out as yet.
%%%%%%%%%%%%%%
\section*{Note Added:}
\noindent After this letter was accepted for publication, we became aware of
the work of Girvin and MacDonald \cite{girvin}, where they 
showed that the gauge-transformed Laughlin
wave-function [ eq. (7) of their paper] shows off-diagonal long-range order.
It then immediately follows that the Calogero-Sutherland ground state
wave-function in two dimensions as given by \eq{grst}  [which is identical to
eq. (7) of \cite{girvin}] also exhibits off-diagonal long-range order.
%%%%%%%%%%%%%%%%%%%%%%%%%%%%%%%%%%%%%%%%%%%%%%

\end{document}